\documentclass{pasj00}
\usepackage{natbib}
\usepackage[dvips]{graphics}

\begin{document}
\SetRunningHead{T. Takagi et al.}{Multi-$\lambda$ analysis of AKARI 18\,$\mu$m-selected 
galaxies}
\Received{}%{yyyy/mm/dd}
\Accepted{}%{yyyy/mm/dd}

\title{Multi-wavelength Analysis of 18\,$\mu$m-selected Galaxies in the AKARI/IRC
monitor field towards the North Ecliptic Pole}

%%% begin:list of authors
% Do NOT capitalize all letters in "textsc".
\author{
Toshinobu \textsc{Takagi},\altaffilmark{1} 
    \thanks{E-mail: takagi@ir.isas.jaxa.jp}
Hideo \textsc{Matsuhara},\altaffilmark{1} 
Takehiko \textsc{Wada},\altaffilmark{1} 
   Shinki \textsc{Oyabu},\altaffilmark{1}  \\
Koji  \textsc{Imai},\altaffilmark{1}   
   Chris P. \textsc{Pearson},\altaffilmark{1,2} 
   Hitoshi \textsc{Hanami},\altaffilmark{3}    
   Takashi \textsc{Onaka},\altaffilmark{4}   \\
   Naofumi \textsc{Fujishiro},\altaffilmark{1} 
   Daisuke \textsc{Ishihara},\altaffilmark{4} 
   Yoshifusa \textsc{Ita}, \altaffilmark{1}
   Hirokazu \textsc{Kataza},\altaffilmark{1} \\ 
   Woojung \textsc{Kim},\altaffilmark{1} 
   Toshio \textsc{Matsumoto},\altaffilmark{1}
   Hiroshi \textsc{Murakami},\altaffilmark{1}
   Youichi \textsc{Ohyama},\altaffilmark{1} \\ 
   Itsuki \textsc{Sakon},\altaffilmark{4}  
   Toshihiko \textsc{Tanab\'e},\altaffilmark{5}
   Kazunori \textsc{Uemizu},\altaffilmark{1}
   Munetaka \textsc{Ueno},\altaffilmark{6}  \\
   Hidenori \textsc{Watarai},\altaffilmark{7}  
   Fumihiko \textsc{Usui},\altaffilmark{1}   
   Hyung Mok \textsc{Lee},\altaffilmark{8} 
   Myungshin \textsc{Im},\altaffilmark{8}  \\
   Stephen \textsc{Serjeant},\altaffilmark{9}  
   Richard S. \textsc{Savage},\altaffilmark{10} \\
   Tsutomu \textsc{Tange},\altaffilmark{1} 
   and
   Takao \textsc{Nakagawa}\altaffilmark{1} 
} %
%  \thanks{Example: Present Address is xxxxxxxxxx}}
 \altaffiltext{1}{Institute of Space and Astronautical Science, Japan Aerospace Exploration Agency, \\
   Sagamihara, Kanagawa 229-8510 }  
 \altaffiltext{2}{ISO Data Centre, ESA, Villafranca del Castillo, Madrid, Spain.}  
  \altaffiltext{3}{Physics Section, Faculty of Humanities and Social Sciences, \\ Iwate University, 
Morioka, 020-8550}  
 \altaffiltext{4}{Department of Astronomy, School of Science, 
 University of Tokyo, \\ Bunkyo-ku, Tokyo 113-0033 }
% \altaffiltext{2}{Graduate School of Science, Nagoya University, Chikusa-ku, Nagoya, 464-8602  }
% \altaffiltext{3}{University of Kent, Canterbury, Kent CT2 7NR, UK }
 \altaffiltext{5}{Institute of Astronomy, University of Tokyo, Mitaka,Tokyo 181-0015}
 \altaffiltext{6}{Department of Earth Science and Astronomy, Graduate School of Arts and Sciences,\\
The University of Tokyo, Meguro-ku,Tokyo 153-8902}
 \altaffiltext{7}{ALOS Project, Japan Aerospace Exploration Agency, Tsukuba, Ibaraki 305-8505 }
 \altaffiltext{8}{Department of Physics \& Astronomy, FPRD, Seoul National University, \\
Shillim-Dong, Kwanak-Gu, Seoul 151-742, Korea}
 \altaffiltext{9}{Astrophysics Group, Department of Physics, \\ The Open University, Milton Keynes, MK7 6AA, UK}
  \altaffiltext{10}{Astronomy Centre, University of Sussex, Falmer, Brighton BN1 9QJ, UK}
% \altaffiltext{6}{Laboratoire d'Astrophysique de Marseille, Traverse du Siphon, BP8 13376 \\
%    Marseille Cedex 12, France}
% \altaffiltext{7}{National Astronomical Observatory of Japan, Mitaka, Tokyo 181-8588}

 %\altaffiltext{10}{Kyung Hee University, 1 Seocheon-dong, Giheung-gu, Yongin-si
 %   Gyeonggi-do 446-701, Korea} 

% \altaffiltext{13}{Astronomical Institute, Tohoku University, Aoba-ku, Sendai 980-8578 }
% \altaffiltext{12}{Institute for Astronomy, University of Hawaii, 2680 Woodlawn Drive, Honolulu, HI 96822, USA}
% \altaffiltext{13}{Carnegie Mellon University, 5000 Forbes Avenue, Pittsburgh, PA 15213, USA}
%%% end:list of authors

%%% Please use the following style in case that sorting by 
%%% affilation is impossible. 
%
% \author{%
%   D-Firstname \textsc{D-Familyname}\altaffilmark{1}
%   E-Firstname \textsc{E-Familyname}\altaffilmark{1,2}
%   and
%   F-Firstname \textsc{F-Familyname}\altaffilmark{2}}
% \altaffiltext{1}{Address of Institute}
% \email{ddddd@xxx.xxx.xx.xx}
% \email{eeeee@xxx.xxx.xx.xx}
% \altaffiltext{2}{Address of Institute}

%% `\KeyWords{}' always has to be placed before `\maketitle'.
%Do NOT move this preamble from here!

\maketitle

\begin{abstract}
We present an initial analysis of AKARI 18\,$\mu$m-selected galaxies using 
all 9 photometric bands at 2 -- 24\,$\mu$m available in the InfraRed Camera (IRC), 
in order to demonstrate 
new capabilities of AKARI cosmological surveys with this unprecedented wavelength 
coverage at mid-infrared bands. We detected 72 sources at 18\,$\mu$m in an area of 
50.2 arcmin$^2$ in the AKARI/IRC monitor field towards the North Ecliptic Pole (NEP). 
From this sample, 25 galaxies with probable redshifts $z\gtrsim 0.5$ are selected with a single colour cut 
($N2-N3>0.1$) for a detailed SED analysis with ground-based $BVRi'z'JK$ data.
Using an SED radiative transfer model of starbursts covering the wavelength 
range UV -- submm, we derive photometric redshifts from the optical-MIR SEDs 
of 18\,$\mu$m-selected galaxies. From the best-fit SED models, we show that 
the IRC all-band photometry is capable of tracing the steep rise in flux at the 
blue side of the PAH 6.2\,$\mu$m emission feature. 
This indicates that the IRC all-band photometry
is useful to constrain the redshift of infrared galaxies, specifically for 
dusty galaxies with a less prominent 4000\,\AA\ break. 
Also, we find that the flux dip between 
the PAH 7.7 and 11.2\,$\mu$m emission feature is recognizable 
in the observed SEDs of galaxies at $z\sim1$. 
By using such a colour anomaly due to the PAH and silicate 
absorption features, unique samples of ULIRGs 
at $z\sim1$, `silicate-break' galaxies, can be constructed from large cosmological 
surveys of AKARI towards the NEP, i.e.\ the NEP-Deep and NEP-Wide survey. 
For AGN candidates selected with an IRC colour-colour diagram  
($N2-N3$ vs.\ $N3-S7$), we find excess emission in more than 
two MIR bands, compared to the best-fit starburst SED model, which 
suggests the presence of hot dust around the central black hole. 
This pilot study suggests the possibility of detecting many interesting galaxy properties 
in the NEP-Deep and Wide surveys, such as a systematic difference in SEDs 
between high- and low-$z$ ULIRGs, and a large variation of the PAH inter-band 
strength ratio in galaxies at high redshifts. 
\end{abstract}

 %Sensitivity
 
 %Earth shine  
 
\section{Introduction}

The universe is a much more luminous place than seen in the optical, 
so it has been revealed since the launch of the first infrared astronomical satellite, 
IRAS \citep{1984ApJ...278L...1N}. 
Subsequent infrared space missions, i.e.\ ISO \citep{1996A&A...315L..27K} 
and Spitzer \citep{2004ApJS..154....1W} have led to enormous advances in our 
knowledge on star-formation activity, specifically in the distant universe 
\citep[e.g.][and references therein]{2000ARA&A..38..761G,2006astro.ph.10897F}. 
The last mission, AKARI was launched on 22nd February 2006 JST \citep{murakami2007}
and is functioning very well in orbit at the time of writing. 

The deepest extragalactic surveys for infrared luminous galaxies have been carried out 
at 15\,$\mu$m with ISO and 24\,$\mu$m with Spitzer, due to the fact that the confusion 
of point sources and of background cirrus emission is more serious in longer 
wavelengths. A key result from ISO surveys may be summarised in the differential 
number counts at 15\,$\mu$m \citep[see e.g.\ Figure 14 in][]{2000ARA&A..38..761G}, 
which have a characteristic bump at around 300\,$\mu$Jy. 
With these number counts, pure density evolution models are clearly 
rejected, and strong luminosity evolution with $\sim (1+z)^4$ is suggested 
at $0<z<1$ \citep{2000ApJ...541..134X,2001MNRAS.325.1511P,
2001A&A...378....1F,2001ApJ...556..562C}. 
Recently, this strong luminosity evolution at $0<z<1$ has been confirmed 
by Spitzer with much greater statistical significance 
\citep{2005ApJ...630...82P,2006MNRAS.370.1159B,2007astro.ph..1283C}. 

Multi-wavelength surveys at {\it both} 15 and 24\,$\mu$m are not just redundant 
observations, but reveal the problems in our understanding of 
distant infrared galaxies. 
Although the differential number counts at 24\,$\mu$m are found to be similar in shape 
to the 15\,$\mu$m counts, it is not straightforward to explain the 24\,$\mu$m counts 
with the phenomenological models of galaxy counts constrained to fit 
the multi-wavelength counts 
prior to the Spitzer results. The peak in the 24\,$\mu$m differential counts is found at 
200 -- 300\,$\mu$Jy (in $dN/dS\times S^{2.5}$) which is fainter than expected from 
models of galaxy counts. The resolution of this problem has not yet been achieved 
\citep[see however][]{2004ApJS..154..112L,2004ApJS..154...80C,2005MNRAS.358.1417P}. 
Since at 15 and 24\,$\mu$m we observe the PAH emission features of galaxies at 
$z=$1 -- 3, we need SED templates with realistic PAH emission features as a function 
of luminosity for galaxies at $z>1$. Unfortunately, such SED templates are not yet 
available. 

With the InfraRed Spectrograph (IRS) onboard Spitzer, MIR features of high-$z$ galaxies, 
such as the PAH emission feature and silicate absorption feature, become accessible 
\citep{
2005ApJ...622L.105H,
2005ApJ...625L..83L,
2005ApJ...628..604Y,
2006astro.ph..8456H,
2006ApJ...651..101W,
2007ApJ...655L..65M,
2007astro.ph..1409T},
although the observed samples are still very limited. The highest redshift PAH detection has 
been reported in an infrared-luminous Lyman break galaxy at $z=3.01$ \citep{2006astro.ph..8456H}, 
which is likely to be a hyperluminous infrared galaxy.  Currently, the prime high-$z$ targets 
for IRS seem to be submm galaxies 
\citep{2005ApJ...625L..83L,2007ApJ...655L..65M,2006astro.ph.12105P}. 
It is rather surprising that PAH emission features are detected in most 
submm galaxies, 
given that they are typically hyperluminous ($L_{IR} > 10^{13} L_\odot$) galaxies which are 
in the local universe generally AGN-dominated galaxies. There is already a hint that 
the MIR properties of high-$z$ galaxies are systematically different to those in 
the local universe. More systematic studies on this issue are clearly needed. 

With Spitzer surveys, the majority of 
Hyper- and Ultra-Luminous InfraRed Galaxy (Hy/ULIRGs)\footnote{[Hy,U]LIRGs defined 
with $\log L_\mathrm{IR}/L_\odot =$[$>13$, 12 -- 13], where $L_\mathrm{IR}$ is the luminosity 
in the wavelength rage from 8 to 1000\,$\mu$m.} targets at high redshifts may 
be derived from the 24\,$\mu$m source catalogues. 
This may result in incomplete samples of ULIRGs specifically at $z\sim 1.5$, 
at which ULIRGs with strong silicate absorption become very faint in the 24\,$\mu$m band 
\citep{2005MNRAS.357..165T,2007ApJ...656..148A}. \cite{2006ApJ...637..727C} 
obtained a bimodal redshift distribution of 24\,$\mu$m sources as a result of the 
deficit of galaxies at $z\sim 1.5$ \citep[see also][]{2006astro.ph.10897F}. 
More comprehensive 
wavelength coverage at MIR bands should be useful to produce complete 
samples of ULIRGs at $1<z<2$. We point out below that the AKARI multi-wavelength 
surveys are best suited for this purpose, thanks to their 
superb wavelength coverage in the NIR -- MIR. 

%In recent extragalactic surveys with Spitzer, such as GOODS\footnote{the Spitzer GOODS 
%website: \tt http://data.spitzer.caltech.edu/popular/goods/}, 
%SWIRE \citep{2004ApJS..154...54L}, First Look Survey \citep{2006AJ....131.2859F}, 
%extended GOODS field survey \citep{2005ApJ...630...82P,2007astro.ph..1283C}, and 
%other GTO surveys

%Systematic difference in physical properties between low-z and high-z infrared galaxies \\
%* less extinction for a given luminosity (Reddy et al 2006; Takagi et al. 2004) \\
%* Lower dust temperature at high-z (Chapman et al. 2005; Yang et al. 2007) \\
%* PAH in Hyper - Ultra IRGs (Huang et al. 2006 and other IRS papers) \\

%Difference between 15um and 24um population? \\
%* SED modification in Papovich et al (2004) \\
%* Luminosity of dominant population is a function of redshift in downsizing 
%way (Franceschini et al. 2006) \\
%* Multi-wavelength approach is actually 

The most unique scientific programme of AKARI is the all-sky survey, the first major update of IRAS. 
Besides this, another uniqueness of AKARI can be found in surveys of 
selected areas with unprecedented wavelength coverage in the 
NIR -- MIR, with 9 photometric bands from 2 to 24\,$\mu$m using the InfraRed Camera 
\citep[IRC;][]{onaka2007}. 
With the IRC, two main extragalactic surveys have been designed and conducted 
around the North Ecliptic Pole (NEP), a deep $\sim 0.5$\,deg$^2$ survey (the NEP-Deep survey) 
and a shallow 6.2\,deg$^2$ survey (the NEP-Wide survey).
See \cite{2006PASJ...58..673M} for details 
on the survey design of the NEP-Deep and NEP-Wide surveys. 

With this comprehensive wavelength coverage, one clear benefit for extragalactic 
MIR surveys is that a more reliable selection for high-$z$ ULIRGs becomes possible, 
as noted above. 
In a flux limited sample with Spitzer MIPS 24\,$\mu$m, it is likely that 
some fraction of ULIRGs at $z\sim 1.5$ are missed, owing to redshifted silicate 
absorption features. 
With IRC, such ULIRGs would be detected at 15 and/or 18\,$\mu$m. 
These ULIRGs may be called 24\,$\mu$m dropouts or `silicate-break' 
galaxies \citep{2005MNRAS.357..165T}. 
This capability would be very useful to make a complete sample of 
MIR-bright ULIRGs for follow-up observations with MIR spectroscopy. 
Furthermore, the resulting well-sampled SEDs of MIR-selected galaxies allow us to study 
the nature of distant infrared galaxies in detail with the help of SED models, including 
the photometric estimation of redshifts from MIR features.

As a pilot study for a full analysis of the NEP survey, 
here we report the first insight on AKARI 18\,$\mu$m-selected distant infrared galaxies 
seen by all 9 photometric bands available in the IRC. 
\cite{lee2007} provide similar study for an AKARI 11\,$\mu$m-selected sample 
with 6 photometric bands between 2 and 11\,$\mu$m. 
\cite{maruma2007} report the optical properties of an AKARI 15\,$\mu$m-selected sample 
taken from \cite{wada2007} in which results of a deep 15\,$\mu$m survey during the 
performance verification phase are presented. 
In section \ref{obs}, we describe our data and reduction method using the IRC monitor 
observations. For detailed 
analysis of 18\,$\mu$m-selected galaxies, we focus on galaxies likely to lie at $z \gtrsim 0.5$, 
i.e.\ the main targets of interests in the NEP survey. Then, we analyse these galaxies with the 
SED radiative transfer model of starbursts from 
\cite{2003PASJ...55..385T,2003MNRAS.340..813T}. 
In sections \ref{sample} and \ref{fitting}, we give our selection method for galaxies 
at $z\gtrsim 0.5$ and the SED fitting method, respectively. Results are described in 
section \ref{results}. In section \ref{discussion}, we attempt to identify silicate-break 
galaxies in our sample. We summarise our study in section \ref{summary}. 
Throughout this paper, we adopt the cosmology of 
$\Omega_m =0.3$, $\Omega_\Lambda =0.7$ and 
$H_0 =70$$\,$km$\,$sec$^{-1}$$\,$Mpc$^{-1}$.
All magnitudes are given in the AB system, unless 
otherwise explicitly noted.

\section{Observations and data reduction} \label{obs}
\subsection{Data from AKARI}
We have used the IRC monitor observations near the NEP with a field centre of 
$\alpha = 17^\mathrm{h}55^\mathrm{m}24^\mathrm{s}$, $\delta = +66^\circ 37' 32''$ 
(J2000), covering one field of view ($\sim 10'\times10'$) by all 9 IRC photometric bands, 
i.e.\ $N2,\ N3,\ N4,\ S7,\ S9W,\ S11,\ L15,\ L18W$ and $L24$. 
This field is regularly monitored with the IRC, in order to check the stability 
of the detectors. \cite{wada2007} report a deep survey of this field 
at 3 and 7\,$\mu$m, which was conducted during the performance verification phase. 
The observations were done with the Astronomical Observation 
Template (AOT) three filter mode, IRC03 (see Onaka et al. 2007 for details). 
During one pointed observation, three IRC channels, NIR, MIR-S and MIR-L 
can operate simultaneously, covering different wavelength ranges. 
The NIR and MIR-S channels share the same field of view, while the MIR-L channel 
observes a region $\sim 20'$ away from the field centre of NIR/MIR-S. 
In this paper, we use the data from 17 NIR/MIR-S pointed observations 
conducted between 19th June and 4th December 2006, and 5 MIR-L 
pointed observations between 28th June and 30th July 2006. 
We summarise the observation parameters in Table \ref{tab:obs}. 

\subsection{Reduction}
We reduced the data with the IRAF-based 
IRC pipeline software\footnote{Version 061218} (Onaka et al. 2007), in order to 
remove and correct the basic instrumental effects: dark subtraction, linearity 
correction, distortion correction and flat fielding. With the AOT IRC03, there is 
a sufficient number of frames per band (6 -- 9 frames) during one pointed observation to coadd 
the MIR-S data and also the MIR-L data. 
We coadded the MIR-S frames from each pointed observation using the IRC 
pipeline in which sky positions of detected point sources are matched 
to calculate the relative astrometric shift between the frames. We did this after subtracting 
the sky background. For MIR-L, it was found that the sky 
background is too high and sources are too faint to match the 
positions of point sources. 
We used an optional command of the IRC pipeline ({\tt coaddLusingS}) to coadd the MIR-L 
images, in which the relative astrometric shifts of the MIR-L images are derived from those 
of the MIR-S images by adopting the calibration of the relative offset between the 
MIR-L and the MIR-S channels. Thus we obtained coadded 
images of MIR-S and MIR-L for each pointed observation. 

In order to coadd images from different pointed observations, we used the publicly 
available software SWarp\footnote{\tt http://terapix.iap.fr/rubrique.php?id\_rubrique=49}, 
after putting the world coordinate system (WCS) on each image. We put the WCS 
for the MIR-S images using the IRC pipeline command {\tt putwcs} in which 
2MASS sources are identified and the astrometric solution is derived from them. 
For MIR-L, we manually identified three bright sources detected both in the MIR-L 
images and in the final $S11$ image, then determined the WCS for the MIR-L images. 
We used a median coaddition to produce the final images. 

For the NIR channel, we put the WCS into each individual frame using {\tt putwcs}. 
This was successful for all of the NIR frames, although there is a large number of cosmic 
ray hits in each frame. Again, SWarp was used to obtain the final images with 
median coaddition.

%For MIR-L and MIR-S, large scale background is subtracted with median sky. Background is earth shine. In MIR-L some residual remains around the centre of FoV. 

%Image PSF (getting round) with FWHM of ??

%Detection limit from random sky photometry is 45 uJy (1sigma). 

\subsection{Source detection and photometry on the IRC images} \label{phot}
Source detection was performed in a circular area of the final L18W image 
which has a radius of 4\,arcmin covering the deepest region. 
We used SExtractor \citep{1996A&AS..117..393B} in which two or more connected 
$>3\,\sigma$ pixels are identified as a source. This threshold results in the 
detection of sources with the signal-to-noise ratio of $\gtrsim 4.5$. 
This parameter setting for 
SExtractor is a result of our optimization to detect reliable point sources and avoid 
fainter extended sources which are likely to be residuals of the background subtraction. 
We have detected 72 sources in the area of 50.2 arcmin$^2$. 

Since all of the 18\,$\mu$m sources are found to have a counterpart at other IRC bands, 
the reliability of the detected sources is very high. We applied the same source 
detection method for the negative $L18W$ image, and detected only 5 sources. 
From an inspection of the negative image, it is obvious that they are not point sources like 
those detected in the positive image, but rather a result of the negative extended pattern 
caused by the residuals of the sky subtraction. 
We also used the negative $L18W$ image to estimate the typical photometric error. 
On the negative $L18W$ image, we performed aperture photometry at 
random source-free positions (determined in the positive image) 
using the same procedure as with detected sources (see below). 
The distribution of flux measured at random positions has the standard deviation 
of 25\,$\mu$Jy, which represents a typical photometric error. 
% sensitivity.pro at ~irc_surveys/monitor/analysis/  
We note that the faintest source in our catalogue has the flux of 129$\pm27$\,$\mu$Jy. 
% colcol.pro at ~irc_surveys/monitor/initial_paper/ 
% Expected number of sources with this depth is ~150 from 24um survey, 
% such as Papovich et al. 

In the other IRC bands, we identified the counterparts of the detected 18\,$\mu$m sources 
by using the sky coordinates from SExtractor. First, we determined the centroid of the counterparts 
(with {\tt gcntrd.pro} in the IDL astronomy user's library\footnote{\tt 
http://idlastro.gsfc.nasa.gov/contents.html}) using the sky coordinates of the 18\,$\mu$m 
sources as the initial approximation of the centroid. 
%Centroid is computed using a box   %1.5sigma = 0.637*FWHM(3pix)
%of half width equal to $2.8''$ and $4.5"$ for NIR and MIR-S/L, respectively. 
If the newly-determined centroid 
is shifted from the original position by $>3''$, we rejected such a source as 
the counterpart of the 18\,$\mu$m source. We found that this method results in a higher 
success rate of finding the counterparts of 18\,$\mu$m sources than the comparison of 
source catalogs from each IRC band with a certain search radius. 
We visually inspected the result of the cross-identifications in postage stamp images 
(shown in Figures \ref{image1} \& \ref{image2}  for the colour selected sample, see below), 
and verified the correctness of the cross-identifications. 

We performed aperture photometry at the centroid positions of the 18\,$\mu$m sources and 
their counterparts in our IRC images. 
We used aperture radii of 2 and 3 pixels for the NIR and the MIR-S/L images, respectively. 
Note that in the flux calibration of the IRC images with standard stars, aperture 
radii of 10 and 7.5 pixels were used for the NIR and MIR-S/L channels, respectively. 
The aperture corrections used to estimate the total fluxes 
%(encircled by the aperture for the flux calibration)
have been obtained from the empirical PSF derived from bright sources.
For the NIR channel, we used bright sources in our images. 
For the MIR-S and the MIR-L channels, IRAS sources have been used which 
are detected in the 
observations for the NEP-Wide survey.

We found that the measured fluxes in the coadded NIR and MIR-S 
images are systematically lower than those before coaddition. 
This is because we coadded the images, which have elongated PSFs, 
with various position angles using a simple median. 
Using the empirical PSFs, we generated 
`coadded' PSFs adopting the same position angle of the actual observations. 
The correction factors for this effect were estimated from aperture photometry 
of PSFs before and after coaddition. The correction factors are found to be 
5\% for $N2$, 7\% for $N3$ and $N4$, and 3 \% for the MIR-S channel. 

We have not applied colour corrections for the IRC photometry  at this initial stage of 
data reduction activity. The total error on the IRC photometry is estimated about 
10 -- 20 \%. 
%Specifically, the photometric error in N4 (and probably S7) may be as large as 30\% 
%as suggested by the resulting SED of stars. 

\subsection{Optical identification}
The IRC monitor field lies inside the NEP-Deep survey area and is covered by the deep 
$BVRi'z'$ optical observations with Subaru/S-cam, reaching $B=28.4$ to 
$z'=26.2$ AB mag in $5\,\sigma$ (Wada et al. 2007, in prep). 
This field is also covered by KPNO2.1m/FLAMINGOS at 
$J$- and $Ks$-band with a depth of $J=21.6$ and $K_s=19.9$ ($3\,\sigma$) 
in Vega magnitudes \citep[Field-NE in][]{2007astro.ph..2249I}.

Since the astrometric accuracy is better for shorter wavelengths in the IRC bands, 
we adopted the positions from the NIR channel to find the optical counterparts of 
the 18\,$\mu$m sources. We used the 
catalogue of FLAMINGOS $K_s$-band sources with optical Subaru/S-cam identifications
taken from Imai et al. (2007). We identified the nearest $K_s$-band source as the 18\,$\mu$m
source counterpart. The resulting identifications were all visually inspected. 
If the angular separation is larger than $3''$, 
we reject the counterpart without visual inspection. 
We found no $K_s$-band counterparts for three 18\,$\mu$m sources, for which 
we searched for the nearest optical counterpart 
from the Subaru/S-cam catalogue from Wada et al. (2007, in prep). 
A typical angular separation between the IRC positions and the optical positions is found to be 
$0.6''\pm0.2''$. 

\section{Sample selection} \label{sample}
We have detected 72 sources in total at 18\,$\mu$m, which include stars and galaxies at 
various redshifts. Since the main targets of this study and of the NEP surveys are star-forming 
galaxies at moderate -- high redshifts, we adopt a simple colour cut to 
reduce the contamination from low redshift ($z\lesssim 0.5$) galaxies and stars. 

\begin{figure*}
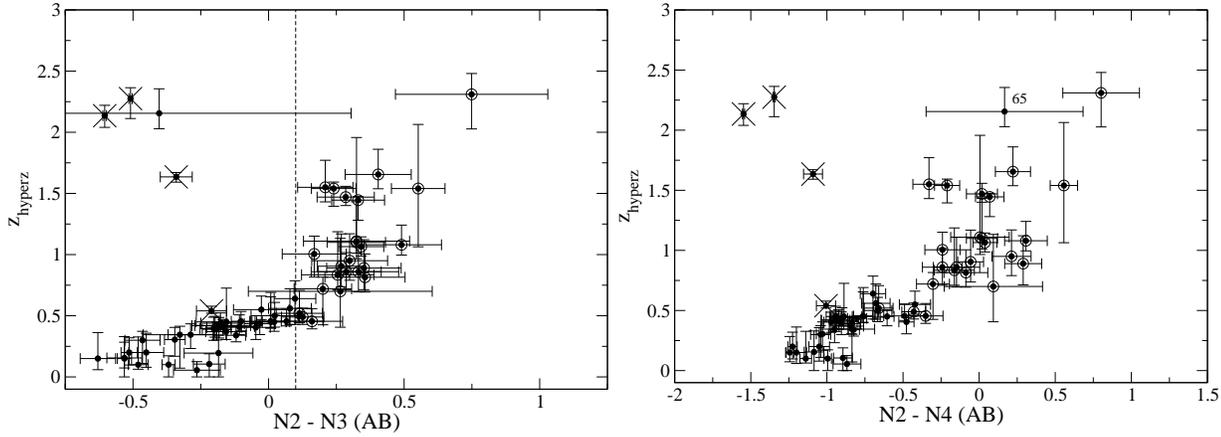
%[b]
\FigureFile(80mm,80mm){figures/n2n3_zphot.eps}
\FigureFile(80mm,80mm){figures/n2n4_zphot.eps}
 \caption{Photometric redshifts derived from the ground-based optical-NIR 
 SED by using {\it hyperz} as a function of the IRC NIR colour, $N2-N3$ and 
 $N2-N4$.  Errors of $z_{hyperz}$ is the 99\% confidence limit. Encircled 
 dots are for galaxies with $N2-N3>0.1$. Most of the galaxies with $N2-N3>0.1$ 
 can be selected with $N2-N4>-0.5$.  Large crosses indicate
 that the best-fit SED model is rejected with the significance of $<1\%$. 
 The tight correlation between photometric redshifts and NIR colours 
 shows that a NIR colour cut can closely mimic a cut with the photometric redshift. 
}
 \label{colz}
\end{figure*}

We did this selection by following \cite{2004ApJS..154...44H} in which $z>0.6$ 
galaxies are selected using only the $K-[3.6\mu$m] colour by utilizing the 1.6\,$\mu$m 
bump. Similar colour selection is possible with the IRC bands, i.e.\ $N2-N3$ and/or 
$N2-N4$. In order to confirm the effectiveness of this single colour selection, we 
derived photometric redshifts of the 18\,$\mu$m sources from the ground-based optical-NIR 
SEDs by using {\it hyperz} \citep{2000A&A...363..476B}. We adopted the SED templates 
of \cite{2003MNRAS.344.1000B} distributed with the source code of {\it hyperz}. 
%Note that these photometric redshifts $z_{hyperz}$ are mainly based on the 4000\,\AA\ break 
%specifically for galaxies at $z\gtrsim 0.3$, i.e.\ a different spectral feature from to that on 
%which $N2-N3$ and $N2-N4$ colours depend.
In Figure \ref{colz}, we plot the derived photometric redshifts against $N2-N3$ and 
$N2-N4$ colours. There is a clear correlation between $z_{hyperz}$ and the 
AKARI NIR colours. This correlation confirms that a rough selection of $z\gtrsim 0.5$ 
galaxies is possible using only the single AKARI NIR colour. We prefer single 
colour cuts to photometric redshifts for the selection of galaxies at $z\gtrsim 0.5$, due to 
their simplicity and availability over the entire field of the NEP survey. 

In the following we focus on the 25 galaxies satisfying the colour cut $N2-N3$$>0.1$ which 
are likely to lie at $z\gtrsim 0.5$. In this sample, we include ID65 which nevertheless 
has $N2-N3<0.1$ with large errors, since its $N2-N4$ is similar to those of the 
other $N2-N3$ selected galaxies and also its overall SED indicates that 
N2 flux is likely to be overestimated. 
As shown in Figure \ref{colz}, a similar selection of galaxies 
is possible with $N2-N4>-0.5$. 

We further divided the 25 sources into two sub-classes, in order to separate AGN candidates 
from normal star-forming galaxies. With Spitzer surveys, the IRAC colour-colour plot of 
$[3.6]-[5.8]$ vs $[4.5]-[8.0]$ has been proposed for use in selecting AGN candidates
\citep{2004ApJS..154..166L}. Similarly, 
\cite{oyabu2007} show that X-ray detected AGNs are found to be systematically red 
in both $N2-N4$ and $N3-S7$ indicating power-law SEDs. Based on their result, 
we adopt a colour cut of $N3-S7>-0.2$ for AGN candidates, which results in 11 AGN
candidates in the 18\,$\mu$m sample with $N2-N3>0.1$. 
In Figure \ref{image1} (\ref{image2}) we show postage stamp images of 14 (11)
$N2-N3$-selected star-forming galaxies (AGN candidates) at $R$- and $K_s$-band 
and 9 IRC bands.

\section{SED fitting method} \label{fitting}
We analyse the SEDs of both 14 normal star-forming galaxies and 11 
AGN candidates using the SED model of StarBUrsts with Radiative Transfer 
\citep[SBURT;][]{2003MNRAS.340..813T,2004MNRAS.355..424T}. Since the 
contribution from AGN is not taken into account in SBURT, it is expected that 
the SEDs of AGN candidates show an excess in the MIR, owing to hot dust around the 
central massive black hole, while the best-fit model for normal star-forming galaxies 
would be largely acceptable. 
%Based on the result of the SED fitting, we derive the 
%physical properties of 18\,$\mu$m-selected star-forming galaxies. 

We find the best-fit SED model from the prepared set of 540 SED models at various 
redshifts with different starburst age (0.01 -- 0.6 Gyr), the 
compactness of the starburst region $0.3 \le \Theta \le 5.0$ and 
the extinction curve (MW, LMC and SMC type)\footnote{These SED models are available at 
\tt http://www.ir.isas.jaxa.jp/\~{}takagi/sedmodel/.}. The compactness of starbursts is 
defined by $r = \Theta (M_*/ 10^9 M_\odot)^\frac{1}{2}$ [kpc], where $r$ and $M_*$ are 
the radius and stellar mass of the starburst region, respectively. 
Note that the attenuation due to dust or 
$A_V$ is a function of starburst age, peaking at $\sim t/t_0$ where $t_0$ is an 
evolutionary time-scale of starbursts taken to be 100 Myr. Also, the more compact the 
starburst region, i.e.\ smaller $\Theta$, the larger the attenuation. 
The adopted SED models are the same as those used in \cite{2004MNRAS.355..424T} 
for the SED analysis of submm galaxies. 

We searched for the best-fit SED model by $\chi^2$-minimization. 
Considering the flux uncertainty with the IRC images and systematic errors between 
ground-based and space photometry, we set minimum flux errors at each 
photometric band, i.e.\ 5\% for optical bands ($BVRi'z'$), 10\% for $J$ and $K_s$, 
15\% for $N2$ and $N3$, 30\% for $N4$ and $S7$, and 
20\% for the other longer IRC bands. 
We have put larger minimum errors on $N4$ and $S7$ bands, since a systematic error 
on calibration is indicated from the SED of stars in our field (see also Lee et al. 2007). 
Since these minimum errors are chosen as the most conservative limit, the derived 
values of $\chi^2$ can be regarded as the lower limits. 

At rest frame UV wavelengths, the SEDs of starbursts could depend on the 
small scale geometry of gas/dust and stars resulting in the leakage of UV photons 
and affecting the resulting SED \citep[c.f.][]{2003MNRAS.340..813T}. Also note that 
the extinction curve is not well understood in starbursts in the UV. 
Therefore, considering these uncertainties in the SED model, we 
quadratically added an additional 20\% error for data at rest-frame UV 
wavelengths, $<$4000\,\AA, in order to reduce the statistical weight of data 
at the rest-frame UV. The adopted flux errors for the SED fitting are 
shown in Figure \ref{sed1} and \ref{sed2}. We reject the SED model if 
the value of $\chi^2$ is significant at $<1$\% level.

\section{Results} \label{results}

\subsection{AGN/starburst diagnostics by the SED fitting}
Figures \ref{sed1} and \ref{sed2} show the best-fit SBURT model for normal star-forming 
galaxies and AGN candidates respectively, including those rejected with 
a large $\chi^2$ value (shown as dotted lines). 
As expected, most of the AGN candidates have power-law-like SEDs (see however ID25 as 
discussed below). According to the $\chi^2$ values, 
SBURT has no acceptable best-fit model for 4 out of 14 star-forming galaxies, while 
7 out of 11 AGN candidates cannot be explained by SBURT. 
For the majority of galaxies with no acceptable SED model, 
the large $\chi^2$ is attributed to the excess of 
observed fluxes in more than 2 MIR bands, compared to the best-fit SED model. 
The detection of the MIR excess, probably due to 
the hot dust component around the central massive black hole, is reliable, 
thanks to the comprehensive wavelength coverage in the MIR. 
All the accepted models for AGN candidates (ID14, 35, 64, and 72) underestimate the 
MIR fluxes as well, but this does not much contribute to the $\chi^2$ value, due to 
the large minimum error (30\%) for $N4$ and $S7$. Thus, if more accurate photometry 
becomes possible for these bands, the rejection rate for AGN candidates could 
reach $\sim$100\%. This result indicates that the simple colour selection of 
AGN candidates with AKARI bands is consistent with the more sophisticated 
diagnostics using the SED fitting. 

%On the other hand, the best-fit models for star-forming 
%galaxies ID11 and 45 are rejected owing to the excess in the MIR bands, which 
%again indicates an additional hot dust component. 

Assuming all of the AGN candidates selected with $N2-N3>0.1$ 
and $N3-S7>-0.2$ are indeed AGN, we estimate the AGN fraction of 18\,$\mu$m-selected 
galaxies to be 15$\pm4$\% (11/72) in number or $\sim$12\% in total flux. 
Although the completeness correction is not yet applied, this fraction is consistent with 
previous MIR surveys \citep[e.g.][]{2002A&A...383..838F,2006ApJ...637..727C}. 
For example, \cite{2002A&A...383..838F} estimate that AGNs contribute 
15$\pm5$\,\% of the total 15\,$\mu$m flux in the Lockman Hole with a survey 
depth of 0.3\,mJy at 15\,$\mu$m. \cite{2004MNRAS.355...97M} similarly find 
an AGN fraction of $\sim 19$\% from Chandra follow up of 0.8 -- 6\,mJy 15\,$\mu$m 
sources from the European Large Area ISO survey.

\subsection{Photometric redshift using MIR features}
MIR photometry, such as Spitzer 24\,$\mu$m band data, are rarely used to derive 
photometric redshifts \citep[see however][]{2005ApJ...630...82P}. 
By using the superb MIR coverage with the IRC photometry, 
it may be possible to constrain the redshifts of distant infrared galaxies
by using the PAH and silicate absorption features. 
In Figure \ref{sed1}, we find that the steep rise in flux corresponding to the blue side of 
the 6.2\,$\mu$m feature is successfully traced by 
the well-sampled IRC photometric bands; the best example of this can be seen in ID10.
This is not possible with Spitzer photometry, which has a large gap between 8 and 24\,$\mu$m.
This new capability of the IRC may be useful for determining photometric redshifts for optically 
faint and featureless galaxies, such as young and dusty starburst galaxies with less prominent 
4000\,\AA\ breaks. 
However, it is still necessary to calibrate the MIR photometric 
redshifts using spectroscopic follow-up observations. 

In Figure \ref{zcomp},  we show the comparison of photometric redshifts derived with 
{\it hyperz} using the ground-based optical-NIR photometry and those with the 
optical-to-IRC bands using SBURT. 
The majority of these photometric redshifts are consistent with one another within the 
99\% confidence limits. Significant discrepancies can be seen in ID14, 24, and 72. 
Among these, ID14 and 72 are AGN candidates, and ID24 is a star-forming galaxy. 
For AGN candidates, photometric redshifts from both {\it hyperz} and SBURT 
would not be reliable, since the 
adopted models do not include an AGN component. 
The star-forming galaxy ID24 has a nearly power-law-like SED in the optical-NIR bands, 
and therefore the photometric redshift using only the optical-NIR SED is likely to be unreliable. 
Photometric redshifts with the IRC bands could be most useful for this type of galaxy.

As suggested by \cite{2005MNRAS.357..165T}, the PAH and silicate absorption features 
of galaxies at $z\sim1$ could be identified using IRC 15, 18 and 24\,$\mu$m photometry. 
Due to the relatively shallow depth of our 24\,$\mu$m image, their `silicate-break' selection 
technique is only partly applicable to our data. We discuss the application 
of this technique for our sample in section \ref{discussion}.

\subsection{The nature of 18\,$\mu$m-selected galaxies} \label{nature}
Here we briefly describe the physical properties of the 18\,$\mu$m-selected galaxies with $N2-N3>0.1$
based on the results of the SED fitting using the SBURT model.
More detailed analyses of AKARI MIR-selected galaxies will be given elsewhere. 

The average 18$\mu$m flux of star-forming galaxies with $N3-S7<-0.2$ is 
240\,$\mu$Jy. In Table \ref{tab:fit}, we summarise the 
derived quantities from the SED fitting of the galaxies with acceptable $\chi^2$ (i.e. not significant 
at $<1$\% level), 
along with the best-fit parameters. 
The average redshift of these galaxies is $\langle z\rangle=1.0$. 
Infrared luminosities range from $10^{11}$ to $3\times 10^{12} L_\odot$, 
% read_fit.pro at goblin:takagi/photoz/akari/
and therefore they are classified as LIRGs or ULIRGs. 
From the SBURT model, a typical stellar mass of these 
(U)LIRGs is derived to be $3\times 10^{10} M_\odot$, including galaxies as massive 
as $\sim 10^{11} M_\odot$.

It is interesting to note that no best-fit SED model shows prominent silicate absorption 
feature, i.e.\ they are less obscured objects, unlike nearby ULIRGs. There may be a 
systematic difference in the dust attenuation between low- and high-$z$ objects with 
the same luminosity. At high redshifts, strong silicate absorption similar to nearby ULIRGs 
\citep[e.g.][]{2007ApJ...656..148A} is found in more luminous objects such as
submm galaxies or MIR-selected galaxies with $>10^{13} L_\odot$ 
\cite[e.g.][]{2005ApJ...622L.105H}. A systematic difference in the SEDs between 
ULIRGs at $z\lesssim1$ and submm galaxies (i.e. Hy/ULIRGs at $z\sim2$) is 
already reported as a systematic difference in dust temperatures
\citep{2005ApJ...622..772C,2007astro.ph..2463Y} and in the UV-submm 
SEDs \citep{2004MNRAS.355..424T}. Also, \cite{2006ApJ...644..792R} claim that 
at a given bolometric luminosity the obscuration of galaxies is generally $\sim10$ 
times smaller at $z\sim2$ than at $z\sim 0$. 
The other possibility is that our results indicate a selection 
effect, since ULIRGs with strong PAH emission should be brighter than 
obscured ULIRGs. 
A systematic difference in the SED of galaxies at different redshifts could be 
one of the most important topics for further work in the NEP survey. \\

\subsubsection*{Cases worthy of comment}

{\bf ID10:} The PAH emission of this galaxy is well sampled with the IRC photometry
given the photometric redshift of $z_\mathrm{phot} = 0.6^{+0.1}_{-0.2}$. With MIR-L bands, the PAH 
11.2\,$\mu$m feature is covered. From the SED fitting, we find that the resulting 
$\chi^2$ is 24.9 and half of this $\chi^2$ value is explained by the $L15$ and $L18W$ 
bands alone, since the model overestimates the observed fluxes. This may indicate that 
the PAH emission of this galaxy has anomalous inter-band strength ratios. 
Certainly, we can put more strong constraints on this inter-band strength 
ratio once the spectroscopic redshift becomes available. 
With the mapping of the nearby spiral galaxy NGC6946 using AKARI/IRC at 7 and 11\,$\mu$m 
and also the MIR spectroscopy, 
\cite{sakon2007} report a strong variation of PAH 7.7/11.2\,$\mu$m ratio, 
depending on the physical properties of interstellar medium. 
In the distant universe, galaxies with anomalous PAH inter-band ratio may exist 
in greater number than expected from the local universe. 

{\bf ID25: } Although this galaxy is 
classified as an AGN candidate, its SED is not characterized by 
a simple power-law. The 1.6\,$\mu$m bump falls in between $N2,\,N3$ and $N4$ bands, 
suggesting that stellar emission dominates in this wavelength range. 
From 7\,$\mu$m, there is a very steep rise in flux towards longer wavelengths 
which cannot be reproduced with the SBURT models. 

{\bf ID26: } This is an extremely red object with $R-K = 4.6$ (6.3 in Vega), and too red for 
the SBURT models within the reasonable parameter range for the SED fitting. Interestingly, 
the SED fitting with {\it hyperz} indicates that this is a young (40 Myr) and heavily 
obscured ($A_V=5$) galaxy at $z=1$.
Also, this galaxy has the bluest $L15-L18W$ colour in our sample, i.e.\ good candidate 
of silicate-break galaxy at $z\sim 1$ (see section \ref{discussion}). 

{\bf ID65: } This galaxy has a very steep power-law SED with the optical-to-IR slope of 
$\alpha \simeq -2.5$ when $f_\nu \propto \nu^\alpha$. This is very close 
to the maximum limit on the steepness for AGN SEDs \citep[e.g.][]{2006ApJ...640..167A}.

\section{Discussion} \label{discussion}
\cite{2005MNRAS.357..165T} proposed a new selection technique of high-$z$ ULIRGs 
by using the AKARI/IRC multi-wavelength survey, in which the colour anomaly in the MIR bands 
due to the silicate absorption feature is used. Galaxies found with this method are called 
silicate-break galaxies. \cite{2005ApJ...634L...1K} adopted this technique 
by using Spitzer 16\,$\mu$m (IRS) and 24\,$\mu$m (MIPS) observations. They found 
36 candidate silicate-break galaxies probably at $1\lesssim z\lesssim 1.8$ in 0.0392\,deg$^2$, 
corresponding $\sim 920$ sources deg$^2$. 

Here we attempt to identify silicate-break galaxies with our data. 
As pointed out by \cite{2005MNRAS.357..165T}, the key wavelength for this selection 
method is 24\,$\mu$m. Unfortunately, our 24$\mu$m image is not deep enough to 
employ the silicate-break technique. Instead, we use $L15$ and $L18W$. 

In Figure \ref{sbg}, we show the colour-colour plot of $N2-N3$ vs.\ $L15-L18W$. 
According to \cite{2005MNRAS.357..165T}, we could select ULIRGs at $z\sim 1$ 
with $L15-L18W \lesssim 0$, but we would suffer from the contamination from
normal spiral galaxies at low redshifts. 
We use a colour cut of $N2-N3>0.1$ again to exclude the 
contamination from low-$z$ galaxies. 

We find that seven 18\,$\mu$m-selected galaxies satisfy $N2-N3>0.1$ and 
$L15-L18W <0$. Five out of seven, ID15, 16, 26, 58 and 60, 
have $N3-S7 < -0.2$, and are likely to be star-forming galaxies. 
Except for ID16 and 58, photometric 
redshifts (from both SBURT and {\it hyperz}) are consistent with the redshift 
expected from the silicate-break selection, i.e.\ $z\sim 1$. 
ID16 and 58 have $z_\mathrm{phot}= 0.6$ and 0.7, respectively. 
These low-$z$ contaminants are 
expected, since the single colour cut $N2-N3>0.1$ is only useful for a rough 
selection of galaxies at $z\gtrsim 0.5$. Such contaminations would be reduced 
if we use a colour selection with 18 and 24\,$\mu$m as suggested by 
\cite{2005MNRAS.357..165T}. 

As noted in section \ref{nature}, no best-fit SED models for silicate-break galaxies 
have strong silicate absorption. So, they are not silicate-break but `PAH'-break galaxies. 
On the other hand, \cite{2005MNRAS.357..165T} show that the SED of a 
silicate-break galaxy selected with the Spitzer IRS peak-up imager can be reproduced 
by an SED model with a strong silicate absorption feature. This galaxy is a submm 
galaxy and therefore more luminous than our 18\,$\mu$m-selected galaxies. 
With deeper and wider MIR-L observations, 
we would find ULIRGs at $z\sim1$ with silicate absorption, i.e.\ true silicate-break galaxies. 
More studies on the nature of silicate-break galaxies will be given elsewhere using 
the main NEP survey.

\section{Summary} \label{summary}
We investigated AKARI 18\,$\mu$m-selected sources by using all nine IRC photometric bands, 
as a pilot study for the main AKARI NEP survey. By using the IRC observations for the 
monitor field, we detected 72 18\,$\mu$m sources from the area of 50.2 arcmin$^2$. %sensitivity! 
In order to select galaxies at $z\gtrsim 0.5$, i.e.\ the main targets for the NEP survey, 
we employed a simple colour selection, $N2-N3 > 0.1$ and obtained a sub-sample of 25. 
AGN candidates in this sub-sample are identified with another colour cut, $N3-S7 > -0.2$, 
which implies a hot dust component. The resulting sample has 14 star-forming 
galaxies at $z\gtrsim 0.5$ and 11 AGN candidates with power-law like SEDs. 

We analysed both star-forming galaxies and AGN candidates with a radiative transfer 
SED model of starbursts, SBURT \citep{2003MNRAS.340..813T}. Since the 
SBURT model does not include an AGN component, the presence of an AGN can be assessed  
by the excess of observed flux at the MIR wavelengths due to hot dust around the 
central massive black hole. We found that our AGN candidates show MIR 
excesses compared to the best-fit SED models, most of which are rejected at the significance 
of $<1\%$ by the resulting large $\chi^2$ value. This means that the simple colour 
selection for AGN candidates is consistent with the more sophisticated diagnostics with 
the SED model. An interesting future project would be SED analysis with more complicated 
models having both starbursts and AGN. 

With the SED fitting, we obtained photometric redshifts using optical-to-MIR SEDs, 
which are largely consistent with ground-based optical-NIR photometric 
redshifts. We demonstrated that the steep rises in flux corresponding to 
the blue-side of the PAH 6.2\,$\mu$m feature and also the flux dip between the PAH 
7.7 and 11.2\,$\mu$m features are recognizable in the resulting SEDs by IRC. 
The IRC all-band photometry is probed to be useful to constrain the redshift 
of infrared galaxies using the MIR spectral features. 

Typically, $N2-N3$-selected 18\,$\mu$m sources are (U)LIRGs at $z\sim1$ with 
masses $\sim 3\times10^{10} M_\odot$. No ULIRG in our sample has strong 
silicate absorption according to the best-fit SED model. This may suggest 
a systematic difference in the nature of ULIRGs at low and high redshifts, or 
may be a simple selection effect, owing to the faintness of ULIRGs with strong silicate 
absorption. We could test these hypotheses with deeper IRC all-band surveys 
such as the NEP-Deep survey.

This pilot study demonstrates well the interesting new capabilities of the AKARI NEP 
survey, compared to the existing infrared surveys by Spitzer. The IRC all-band 
photometry would be a powerful and unique tool to constrain the redshifts of optically 
faint heavily obscured galaxies. 
We may find many candidate examples of infrared galaxies with 
anomalous PAH inter-band strength ratio.
Furthermore, unique samples of ULIRGs at $z\sim 1$ will be constructed with 
the silicate-break selection technique. Such samples could play an important 
role in revealing the nature of high-$z$ ULIRGs, which should provide vital 
clues on  the process of galaxy formation. 

\section*{Acknowledgements} 
We would like to thank all the AKARI team members for their extensive 
efforts, without which this work was not possible at all. TT is grateful to 
N.\ Arimoto for his encouragements and supports. TT would like to thank 
Y.\ Sato for stimulating discussion and useful comments. 
This work is supported by the JSPS grants (grant number 18$\cdot$7747). 
CPP acknowledges support from JSPS while in Japan.
MI was supported by the Korea Science and Engineering Foundation (KOSEF) 
grant funded by the Korea government (MOST), No. R01-2005-000-10610-0.

\onecolumn

%%%%%%%%%%%%%%%%%%%%%%%%%%%%%%%%%%%%%%%
\begin{longtable}{lcccc} 
  \caption{Summary of observations}\label{tab:obs}
  \hline              
Band & $\lambda_\mathrm{ref}^{a)} $ &Integration time & \# of frames \\%& Sensitivity \\ 
           &       [$\mu$m]      & [sec]  &     \\%&    1\,$\sigma$ [$\mu$Jy] \\
\endfirsthead
  \hline
Band & $\lambda_\mathrm{ref}$ &Integration time \\%& \# of Frame & Sensitivity  (1\,$\sigma$, $\mu$Jy)\\ 
\endhead
  \hline
\endfoot
  \hline
\multicolumn{5}{l}{\small $^{a)}$ Reference wavelength}
\endlastfoot
  \hline
N2 &  2.4 &2112.1 & 51  \\ %&  1   \\    % sensitivity.pro with negative image
N3 &  3.2 &2029.3 & 49  \\ %&   1  \\
N4 &  4.1 &1490.9 & 36  \\ %&     1  \\
S7 &  7.0 &2503.6 & 153\\  %&   5  \\
S9W & 9.0&2339.9 & 143 \\ %&  6  \\
S11 & 11.0&1767.2 & 108 \\ %&  9 \\
L15 & 15.0&736.3  & 45\\  %&   25 \\
L18W & 18.0&763.3 & 45 \\  %& 25 \\
L24 & 24.0 &589.1 & 36 \\  %&   80  \\
\end{longtable}

\begin{longtable}{lcccccccc} 
  \caption{Results of the SED fitting with SBURT$^{a)}$}\label{tab:fit}
  \hline              
  ID & red.\,$\chi^{2\,b)}$ & $z_\mathrm{phot}^{c)}$ & Age & $\Theta$ & Ext.$^{d)}$ & 
  $\log L_\mathrm{IR}^{e)}$ & $\log M_*$  & $\log$ SFR$^{f)}$ \\ 
    &     &      & [Gyr] & & & [$L_\odot$] & [$M_\odot$] & [$M_\odot$\,yr$^{-1}$] \\
  \endfirsthead
  \hline
  ID & $red.\,\chi^2$ & $z_\mathrm{phot}$ & Age & $\Theta$ & Ext. & $\log L_\mathrm{IR}$ & $\log M_*$  & $\log$ SFR \\ 
\endhead
  \hline
\endfoot
  \hline
 \multicolumn{8}{l}{\small $^{a)}$Rejected fitting results are not tabulated.} \\  %($\chi^2$ divided by the degree of freedom)}\\
\multicolumn{8}{l}{\small $^{b)}$Reduced $\chi^2$} \\  %($\chi^2$ divided by the degree of freedom)}\\
\multicolumn{8}{l}{\small $^{c)}$Photometric redshifts with errors given at the 99\% level}  \\
\multicolumn{8}{l}{\small $^{d)}$Extinction curve}  \\
\multicolumn{8}{l}{\small $^{e)}$Infrared luminosity with the wavelength range of $\lambda = 8$ -- 1000\,$\mu$m} \\
\multicolumn{8}{l}{\small $^{f)}$Star formation rate} \\
\multicolumn{8}{l}{\small $^{g)}$AGN candidates} \\
\endlastfoot
  \hline
6 &2.14 &$1.3^{+0.5}_{-0.2}$ &0.5 &1.2 &LMC &12.0 &11.0 &2.0 \\
10 &2.07 &$0.6^{+0.1}_{-0.2}$ &0.4 &1.6 &MW &11.4 &10.2 &1.5 \\
14$^{g)}$ &1.39 &$2.0^{+0.3}_{-0.7}$ &0.3 &2.0 &LMC &12.5 &11.1 &2.6 \\
15 &1.16 &$1.1^{+0.2}_{-0.3}$ &0.4 &2.0 &MW &11.7 &10.5 &1.8 \\
16 &1.40 &$0.5^{+0.2}_{-0.1}$ &0.4 &1.4 &LMC &11.1 &9.86 &1.1 \\
24 &1.59 &$1.0^{+0.2}_{-0.1}$ &0.3 &1.4 &SMC &12.1 &10.6 &2.1 \\
35$^{g)}$ &1.16 &$0.6^{+0.1}_{-0.2}$ &0.6 &1.2 &MW &11.1 &10.2 &1.1 \\
39 &1.37 &$1.1^{+0.4}_{-0.2}$ &0.4 &1.6 &MW &11.7 &10.4 &1.7 \\
52 &1.41 &$1.1^{+0.2}_{-0.2}$ &0.2 &2.0 &LMC &12.0 &10.2 &2.1 \\
58 &1.48 &$0.7^{+0.1}_{-0.2}$ &0.4 &1.4 &LMC &11.4 &10.1 &1.4 \\
60 &0.71 &$1.3^{+0.2}_{-0.3}$ &0.2 &2.0 &SMC &12.2 &10.4 &2.3 \\
64$^{g)}$ &0.99 &$1.1^{+0.9}_{-0.3}$ &0.1 &1.6 &SMC &11.9 &9.82 &2.0 \\
68 &1.56 &$0.9^{+0.3}_{-0.2}$ &0.3 &2.2 &MW &11.5 &10.1 &1.6 \\
70 &1.14 &$1.1^{+0.6}_{-0.4}$ &0.3 &2.4 &MW &11.6 &10.1 &1.7 \\
72$^{g)}$ &2.15 &$0.9^{+0.2}_{-0.2}$ &0.1 &1.0 &SMC &12.0 &9.83 &2.1 \\
\end{longtable}

%%%  generated by colcol.pro at ~/irc_surveys/monitor/initial_paper/

%%%  generated by colcol.pro at ~/irc_surveys/monitor/initial_paper/
  \begin{figure}%[b]
\FigureFile(80mm,80mm){figures/n2n3s7_col.eps}
 \caption{$N2-N3$ vs.\ $N3-S7$ colour-colour 
 diagram to classify 18\,$\mu$m-selected 
 galaxies into low-$z$ ($z<0.5$) galaxies, high-$z$ star-forming galaxies 
 and AGN candidates. 
}
 \label{col237}
\end{figure}

%%%  generated by check_position.pro at ~/irc_surveys/monitor/analysis/
\begin{figure*}
\begin{center}
 \resizebox{17cm}{!}{\includegraphics{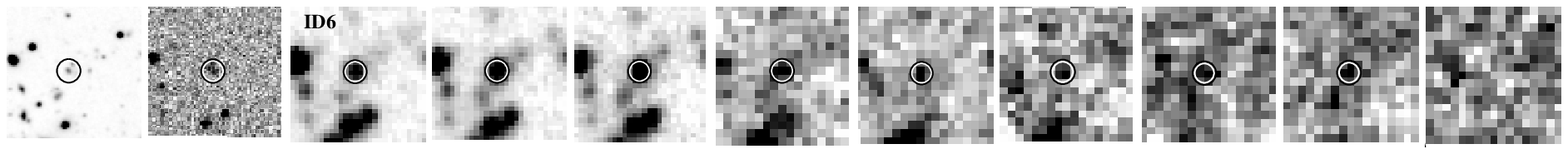}}
 \resizebox{17cm}{!}{\includegraphics{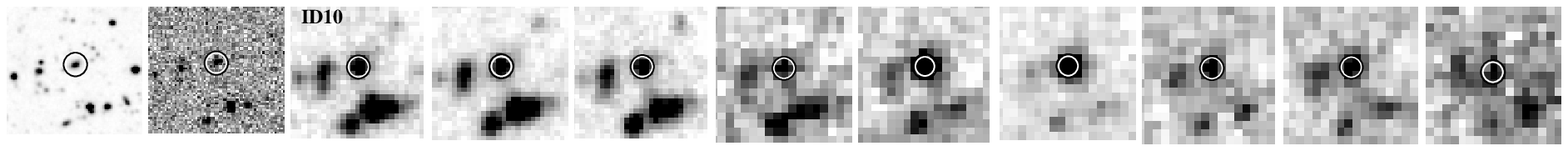}}
\resizebox{17cm}{!}{\includegraphics{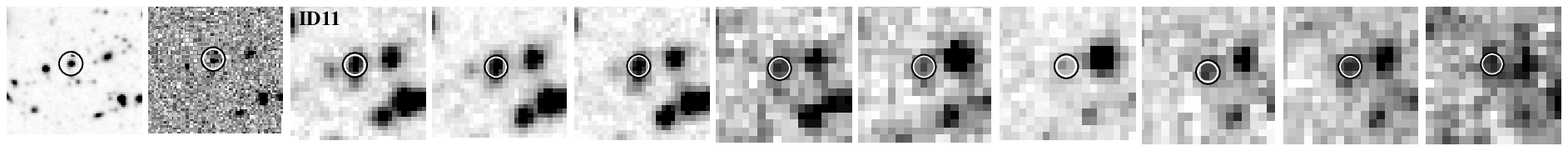}}
\resizebox{17cm}{!}{\includegraphics{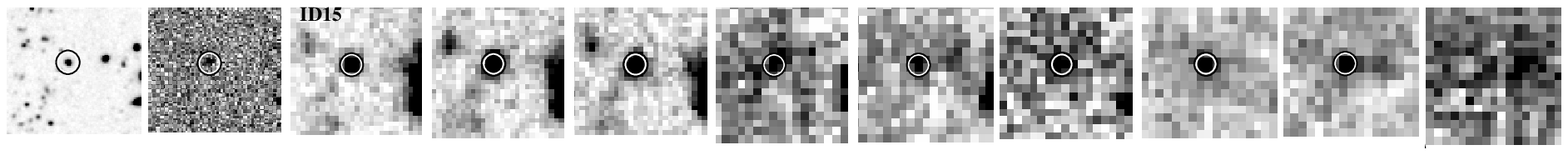}}
\resizebox{17cm}{!}{\includegraphics{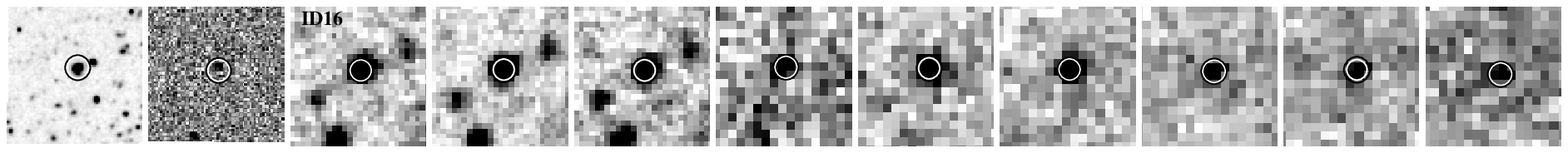}}
\resizebox{17cm}{!}{\includegraphics{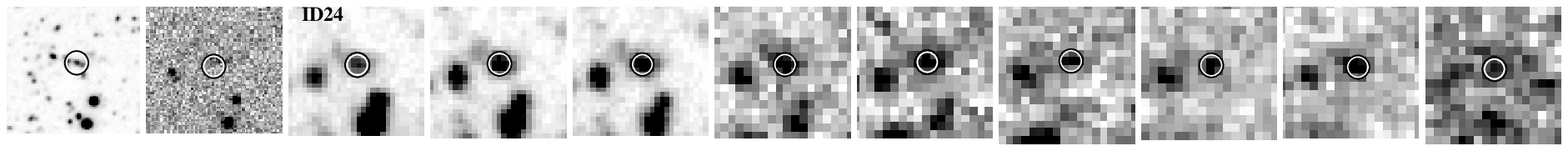}}
\resizebox{17cm}{!}{\includegraphics{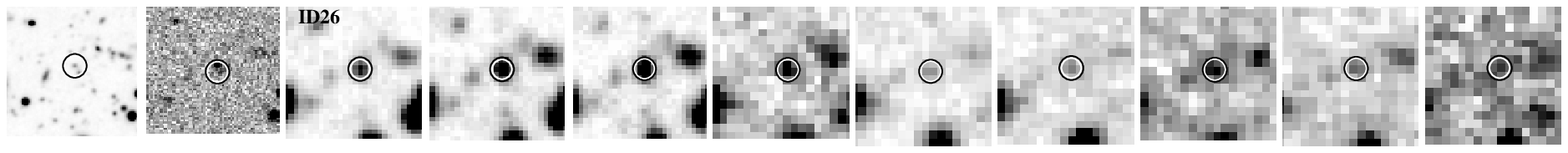}}
\resizebox{17cm}{!}{\includegraphics{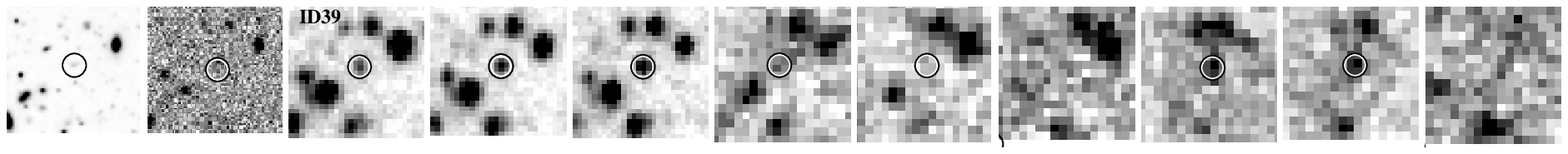}}
\resizebox{17cm}{!}{\includegraphics{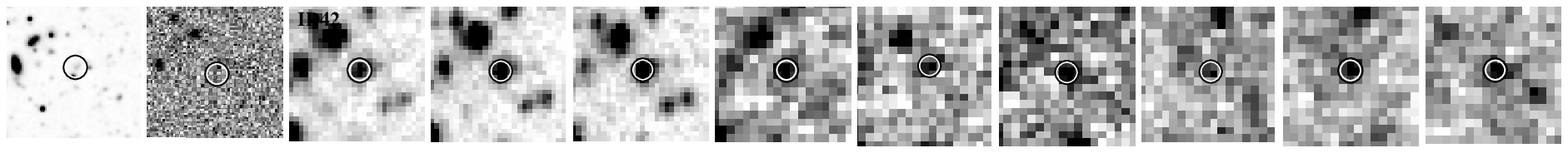}}
\resizebox{17cm}{!}{\includegraphics{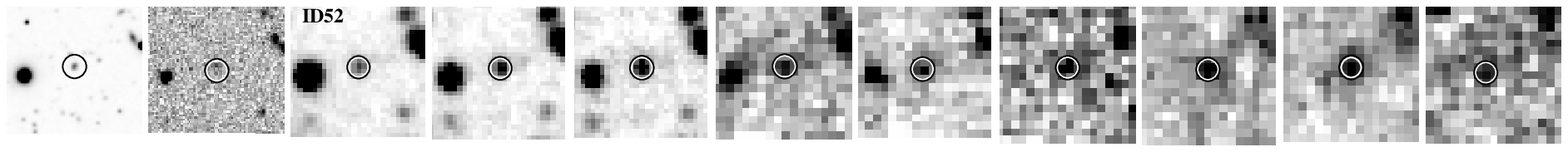}}
\resizebox{17cm}{!}{\includegraphics{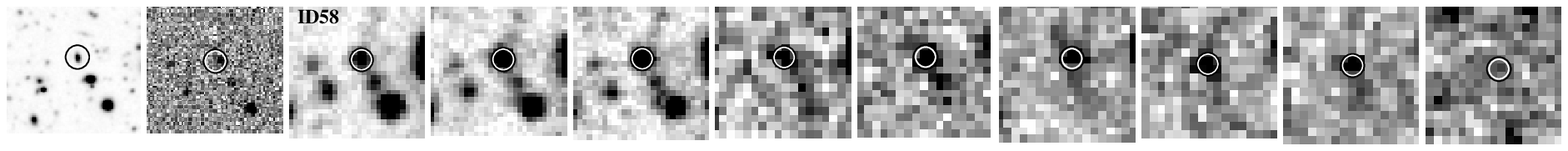}}
\resizebox{17cm}{!}{\includegraphics{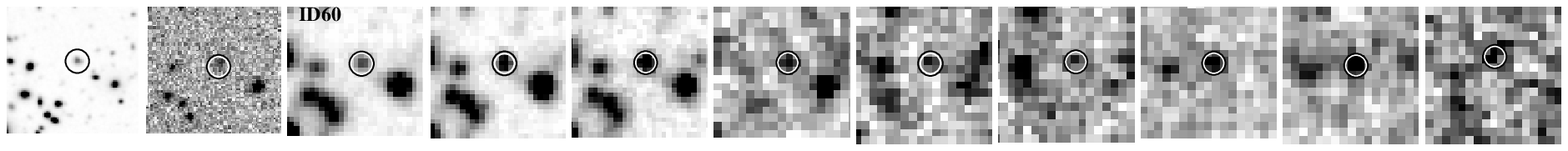}}
\resizebox{17cm}{!}{\includegraphics{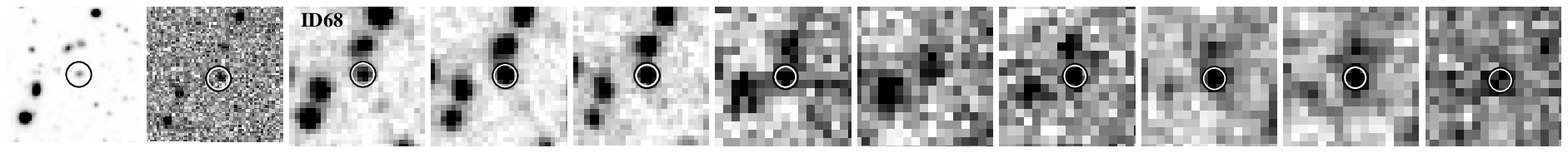}}
\resizebox{17cm}{!}{\includegraphics{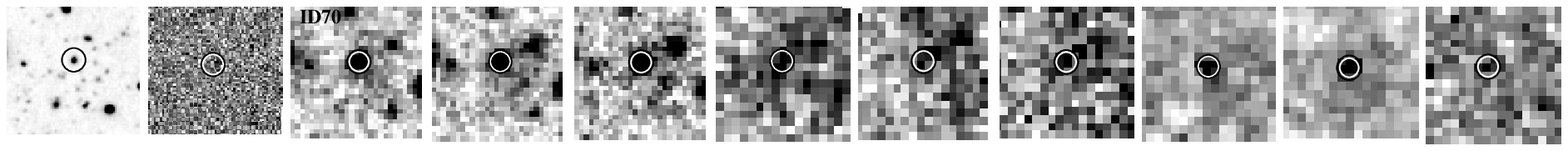}}
 \caption{Postage stamp image of 18\,$\mu$m-selected galaxies with 
 $N2-N3>0.1$ and $N3-S7 < -0.2$, i.e. star-forming galaxies at $z\gtrsim0.5$. 
 From left to right, we show $R$, $K_s$, $N2,\, N3,\, N4,\, S7,S9W,\,S11,\,L15,\,
 L18W$ and $L24$ band images. Identified $L18W$ 
 sources are indicated with black and white circles with the radii of $3.5''$ and $3''$ respectively.
 The north is up and the east is to the left. 
}
 \label{image1}
\end{center}
\end{figure*}

\begin{figure*}
\begin{center}
 \resizebox{17cm}{!}{\includegraphics{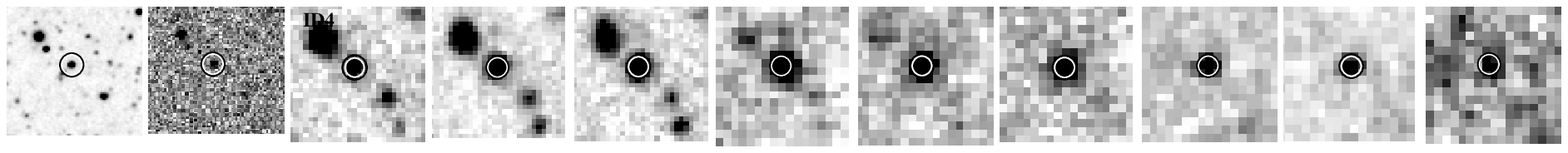}}
 \resizebox{17cm}{!}{\includegraphics{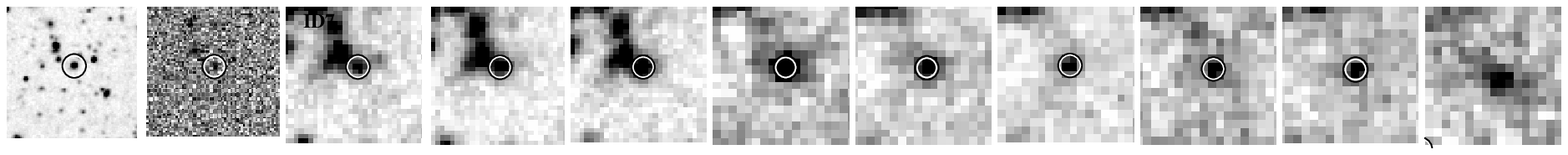}}
\resizebox{17cm}{!}{\includegraphics{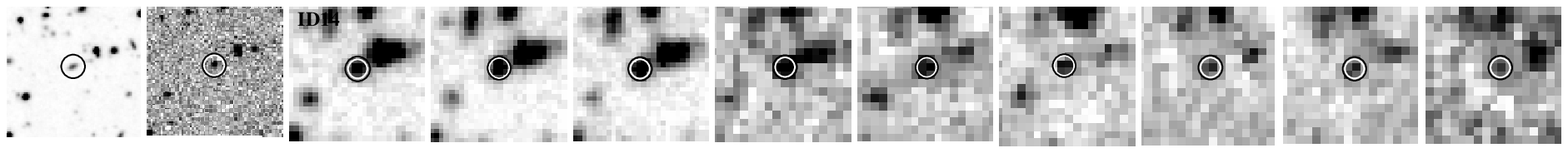}}
\resizebox{17cm}{!}{\includegraphics{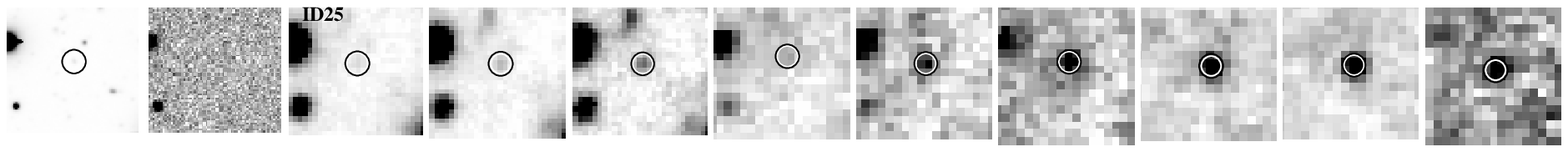}}
\resizebox{17cm}{!}{\includegraphics{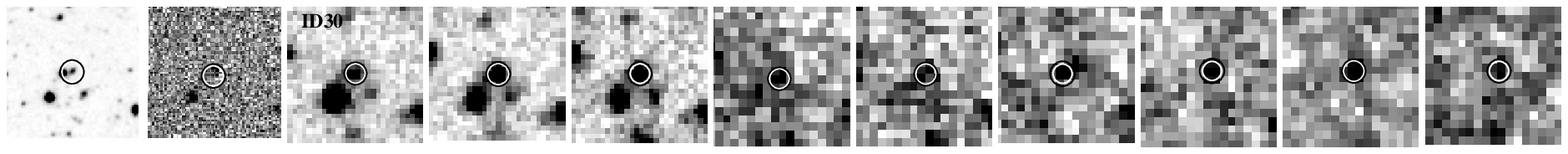}}
\resizebox{17cm}{!}{\includegraphics{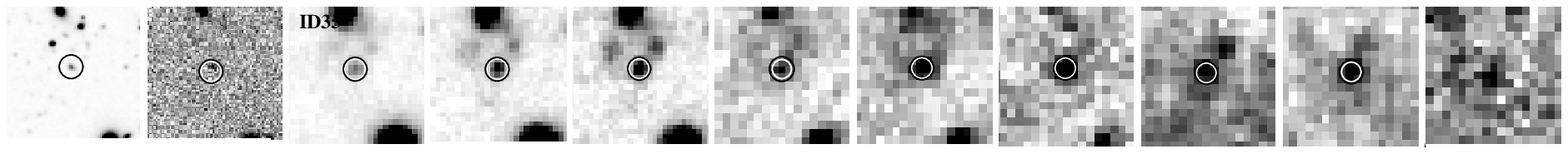}}
\resizebox{17cm}{!}{\includegraphics{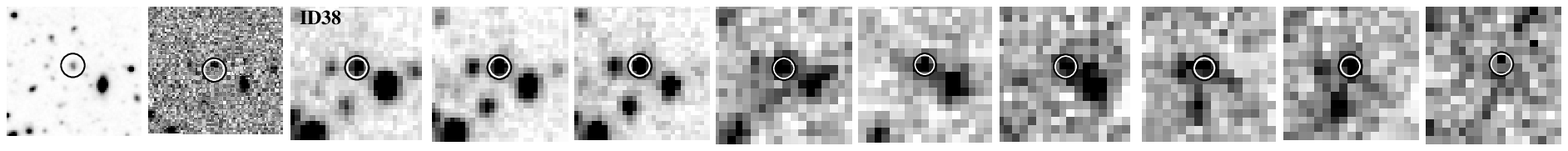}}
\resizebox{17cm}{!}{\includegraphics{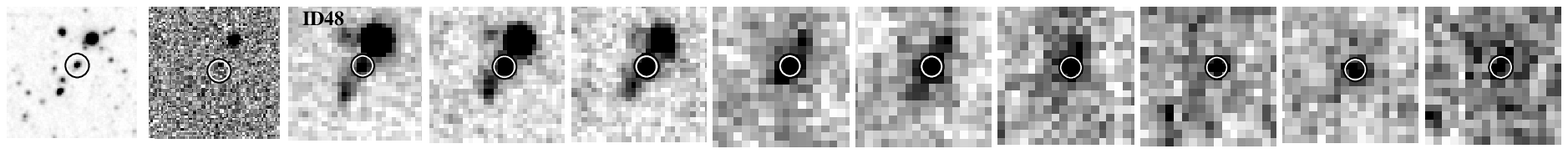}}
\resizebox{17cm}{!}{\includegraphics{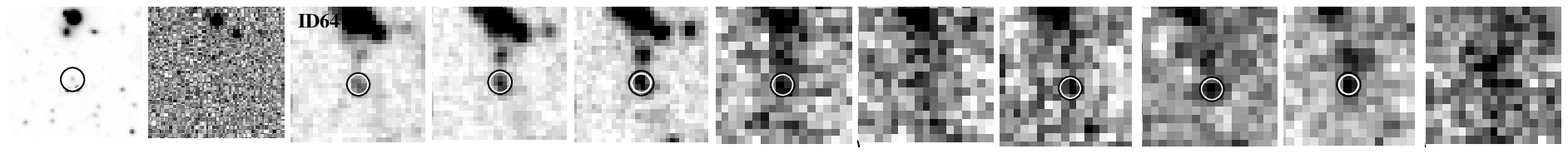}}
\resizebox{17cm}{!}{\includegraphics{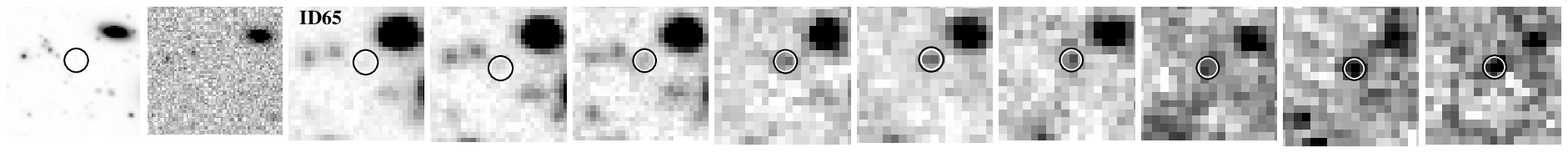}}
\resizebox{17cm}{!}{\includegraphics{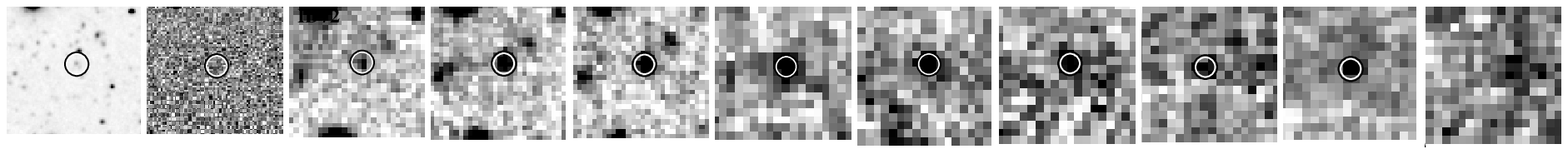}}
%\FigureFile(170mm,!){figures/postage_stamp/l18w_id72.eps}
%\FigureFile(170mm,20mm){figures/postage_stamp/l18w_id72.eps}
 \caption{Same as Figure \ref{image1}, but for galaxies with 
 $N2-N3>0.1$ and $N3-S7 > -0.2$, i.e. AGN candidates. 
}
 \label{image2}
\end{center}
\end{figure*}

  \begin{figure*}
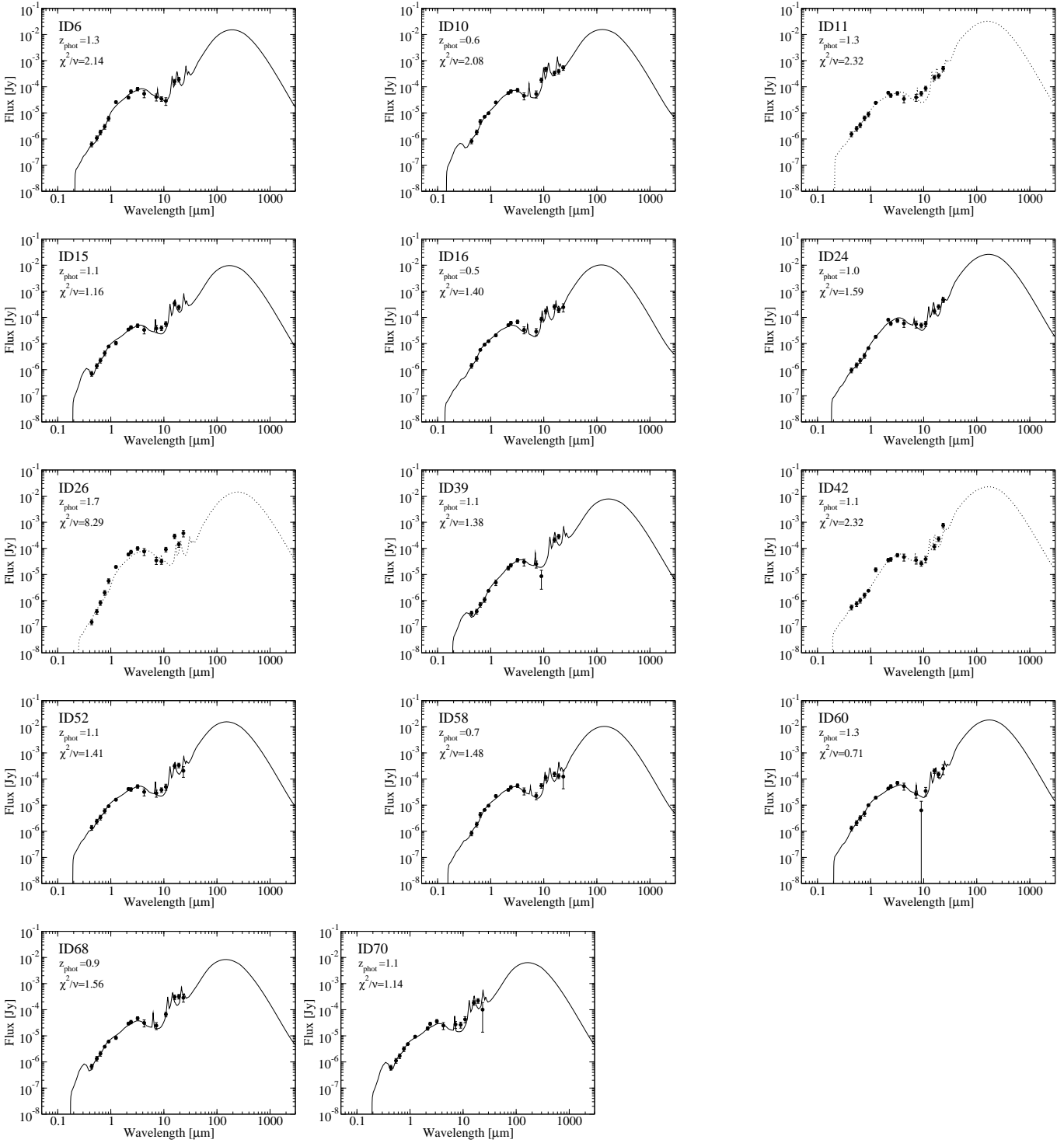
%[b]
\FigureFile(50mm,50mm){figures/sed_id6.eps}
  \FigureFile(50mm,50mm){figures/sed_id10.eps}
    \FigureFile(50mm,50mm){figures/sed_id11.eps}
      \FigureFile(50mm,50mm){figures/sed_id15.eps}
      \FigureFile(50mm,50mm){figures/sed_id16.eps}
      \FigureFile(50mm,50mm){figures/sed_id24.eps}           
      \FigureFile(50mm,50mm){figures/sed_id26.eps}
      \FigureFile(50mm,50mm){figures/sed_id39.eps}
      \FigureFile(50mm,50mm){figures/sed_id42.eps}           
      \FigureFile(50mm,50mm){figures/sed_id52.eps}
      \FigureFile(50mm,50mm){figures/sed_id58.eps}       
      \FigureFile(50mm,50mm){figures/sed_id60.eps}
      \FigureFile(50mm,50mm){figures/sed_id68.eps}
      \FigureFile(50mm,50mm){figures/sed_id70.eps}                                     
 \caption{
Results of SED fitting for 18\,$\mu$m-selected galaxies with 
 $N2-N3>0.1$ and $N3-S7 < -0.2$, i.e. star-forming galaxies at $z\gtrsim0.5$.
 The best-fit SED models rejected with the significance of $<1\%$ are shown in 
 dotted lines. 
}
 \label{sed1}
\end{figure*}

  \begin{figure*}%[b]
 \FigureFile(50mm,50mm){figures/sed_id4.eps}
   \FigureFile(50mm,50mm){figures/sed_id7.eps}
     \FigureFile(50mm,50mm){figures/sed_id14.eps}
       \FigureFile(50mm,50mm){figures/sed_id25.eps}
       \FigureFile(50mm,50mm){figures/sed_id30.eps}
       \FigureFile(50mm,50mm){figures/sed_id35.eps}           
       \FigureFile(50mm,50mm){figures/sed_id38.eps}
       \FigureFile(50mm,50mm){figures/sed_id48.eps}
       \FigureFile(50mm,50mm){figures/sed_id64.eps}   
       \FigureFile(50mm,50mm){figures/sed_id65.eps}        
       \FigureFile(50mm,50mm){figures/sed_id72.eps}
 \caption{Same as Figure \ref{sed1}, but for galaxies with 
 $N2-N3>0.1$ and $N3-S7 > -0.2$, i.e. AGN candidates. 
}
 \label{sed2}
\end{figure*}

  \begin{figure}
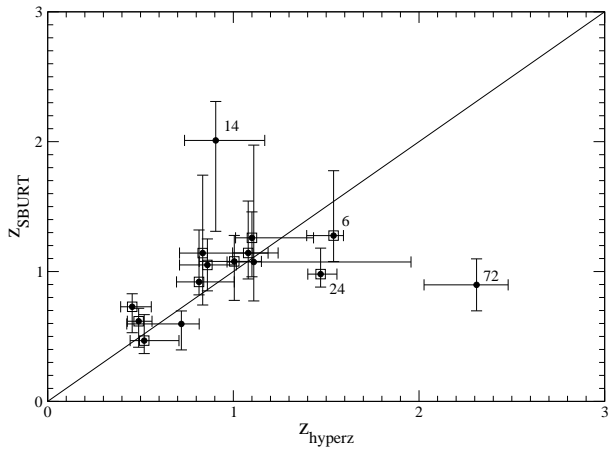
%[b]
    \FigureFile(80mm,80mm){figures/zcomp.eps}
%    \resizebox{10cm}{!}{\includegraphics{figures/zcomp.eps}}
 \caption{Comparison of photometric redshifts derived from 
 {\it hyperz} and SBURT. Errors on photometric redshifts are 
in 99\% confidence limit. The numbers indicate the object ID 
of corresponding data points. Squares indicate star-forming galaxies. 
}
 \label{zcomp}
\end{figure}

  \begin{figure}
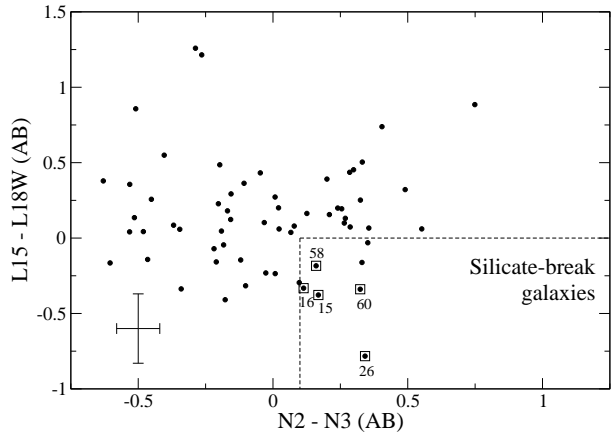
%[b]
    \FigureFile(80mm,80mm){figures/n2n3l15l18_colcol.eps}
 \caption{Colour-colour plot to identify silicate-break galaxies. 
 Squares indicate star-forming galaxies which satisfy selection 
 criteria of silicate-break galaxies. The numbers indicate the object ID 
of corresponding data points. 
}
 \label{sbg}
\end{figure}

\newpage

%  \begin{figure}%[b]
%    \FigureFile(80mm,80mm){figures/bzk_RK.eps}
% \caption{This may not be used in this paper. 
% The numbers indicate the object ID of corresponding data points. 
%}
% \label{bzkrk}
%\end{figure}

\end{document}